\documentclass[twocolumn,pdftex]{revtex4}

\usepackage[pdftex]{graphicx}

\begin{document}

\title{Yakutsk Array Radio Emission Registration Results in the Energy Range of 3$\cdot$10$^{16}$-5$\cdot$10$^{18}$ eV}


\author{I. Petrov}
\author{S. Knurenko}
\author{Z. Petrov}
\author{V. Kozlov}
\author{M. Pravdin}

\affiliation{Yu. G. Shafer Institute of Cosmophysical Research and Aeronomy Siberian Branch of the RAoS, Russia}

\email{igor.petrov@ikfia.sbras.ru}

\begin{abstract}
  This paper presents the set of measurements of ultra-high energy air shower radio emission at frequency 32 MHz in period of 2008-2012. The showers are selected by geomagnetic and azimuth angles and then by the energy in three intervals: 3$\cdot$10$^{16}$ – 3$\cdot$10$^{17}$ eV, 3$\cdot$10$^{17}$ – 6$\cdot$10$^{17}$ eV and 6$\cdot$10$^{17}$ – 5$\cdot$10$^{18}$ eV. In each energy interval average lateral distribution function using mathematically averaged data from antennas with different directions are plotted. In the paper, using experimental data the dependence of radio signal averaged amplitude from geomagnetic angle, the shower axis distance and the energy are determined. Depth of maximum of cosmic ray showers Xmax for the given energy range is evaluated. The evaluation is made according QGSJET model calculations and average lateral distribution function shape.
\end{abstract}


\maketitle

\section{Introduction}

One of the techniques to register ultra-high energy extensive air showers (EAS) is measuring strength of radio pulse by antennas. Unlike traditional techniques, including optic measurements of air shower propagation radio technique can operate in any atmospheric condition except during thunderstorm conditions for whole observation period, which dramatically increases effective time of air showers registration. It is easier to use and much cheaper than other ground detectors in existing air showers array.

	The Yakutsk array measured three components of air shower: the total charged component, the muon component and Cherenkov radiation. From these components using average lateral distribution function (LDF) the integral characteristics of air shower, the total number of charged particles, the total number of muons and full flux of Cherenkov light at the sea level are recovered. All these shower characteristics are used for further model-free air shower energy estimation as shown in [1]. Cherenkov light registered at the sea level moreover is used to recover air shower longitudinal distribution and it characteristics, cascade curve and depth of maximum X$_{max}$ [2][3]. Using this, in future is possible to find a relation between the characteristics of the radio emission and characteristics of the EAS, including slope of the radio emission LDF with depth of maximum, as shown in [4].

\section{Radio Event Selection for Analysis}

For the season 2008-2012 were recorded 600 air shower events with radio emission. Showers energy were above 3$\cdot$10$^{16}$ eV, and zenith angle
$\theta$ $\leq$ 70$^\circ$. For further analysis were selected only 421 showers, appropriate selection criteria of this paper. Therefore, for analysis at the Yakutsk array we use following criteria:
\begin{enumerate}
  \item	The shower selected if ADC prehistory contains radio pulse with amplitude 5 times more than noise level and pulse is localized within time gate equal to delay of “master” from small or large arrays.
  \item Extensive air shower axis must be within perimeter of central array with radius 600 m. Zenith angle $\theta$ $\leq$ 35$^\circ$. Azimuth angle $\phi$ chosen such a way as to exclude influence of polarization effect. That is, the amplitude of the crossed antennas were equal or weren't go beyond limit (3-5) $\%$.
\end{enumerate}

\begin{figure}[t]
  \centering
  \includegraphics[width=0.4\textwidth]{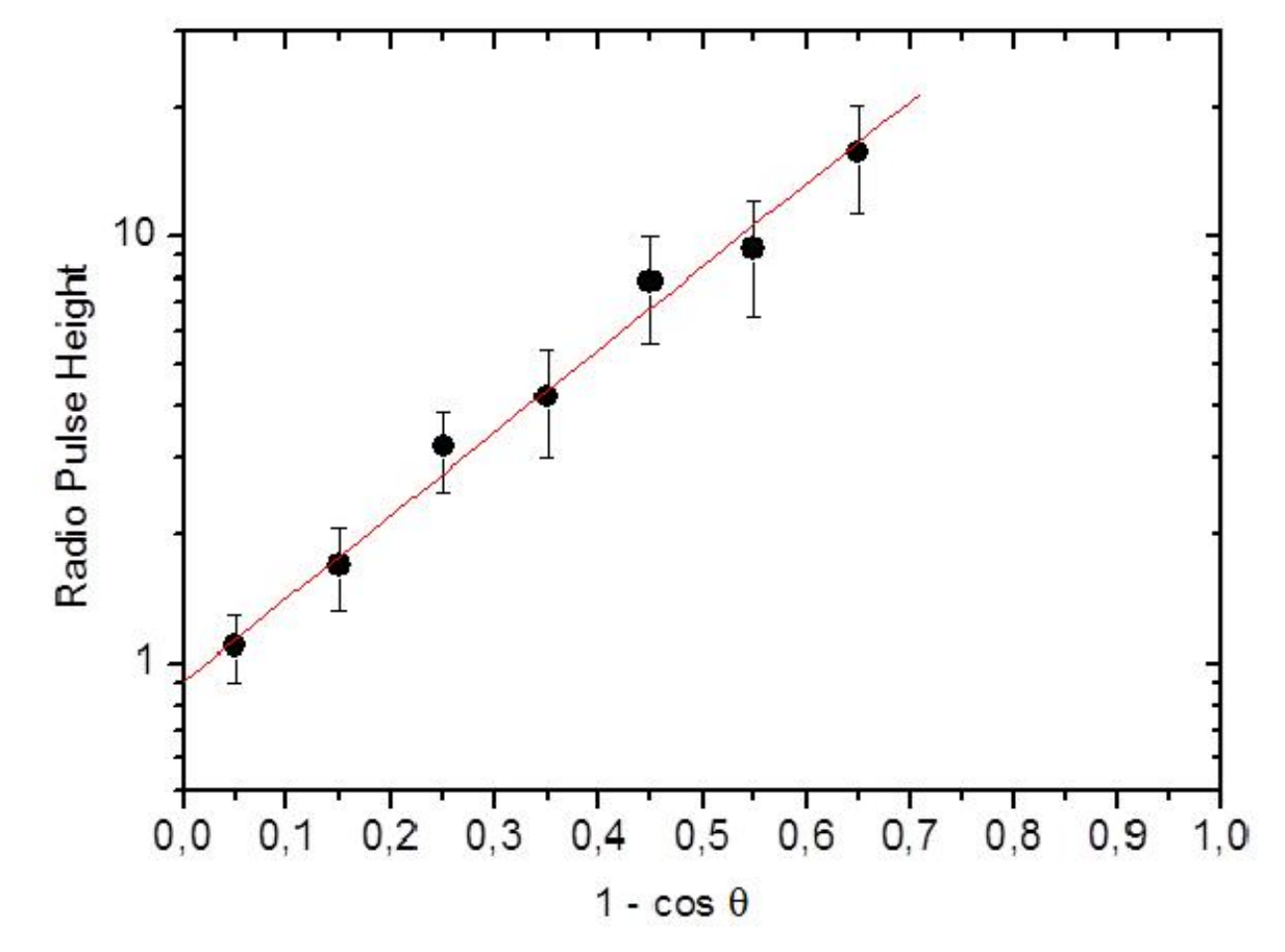}
  \caption{Dependence of maximum amplitude of radio pulse from zenith angle.}
  \label{icrc2013-0181-01}
 \end{figure}

With selected events, we plotted dependence of maximum amplitude of radio pulse from zenith angle (Fig. $\ref{icrc2013-0181-01}$). Approximation curve is given by power function:

\begin{equation}
    \varepsilon(\theta) = (0.81\pm0.25)(1-\cos\theta)^{1.16\pm0.05}
\end{equation}

\begin{figure}[t]
  \centering
  \includegraphics[width=0.4\textwidth]{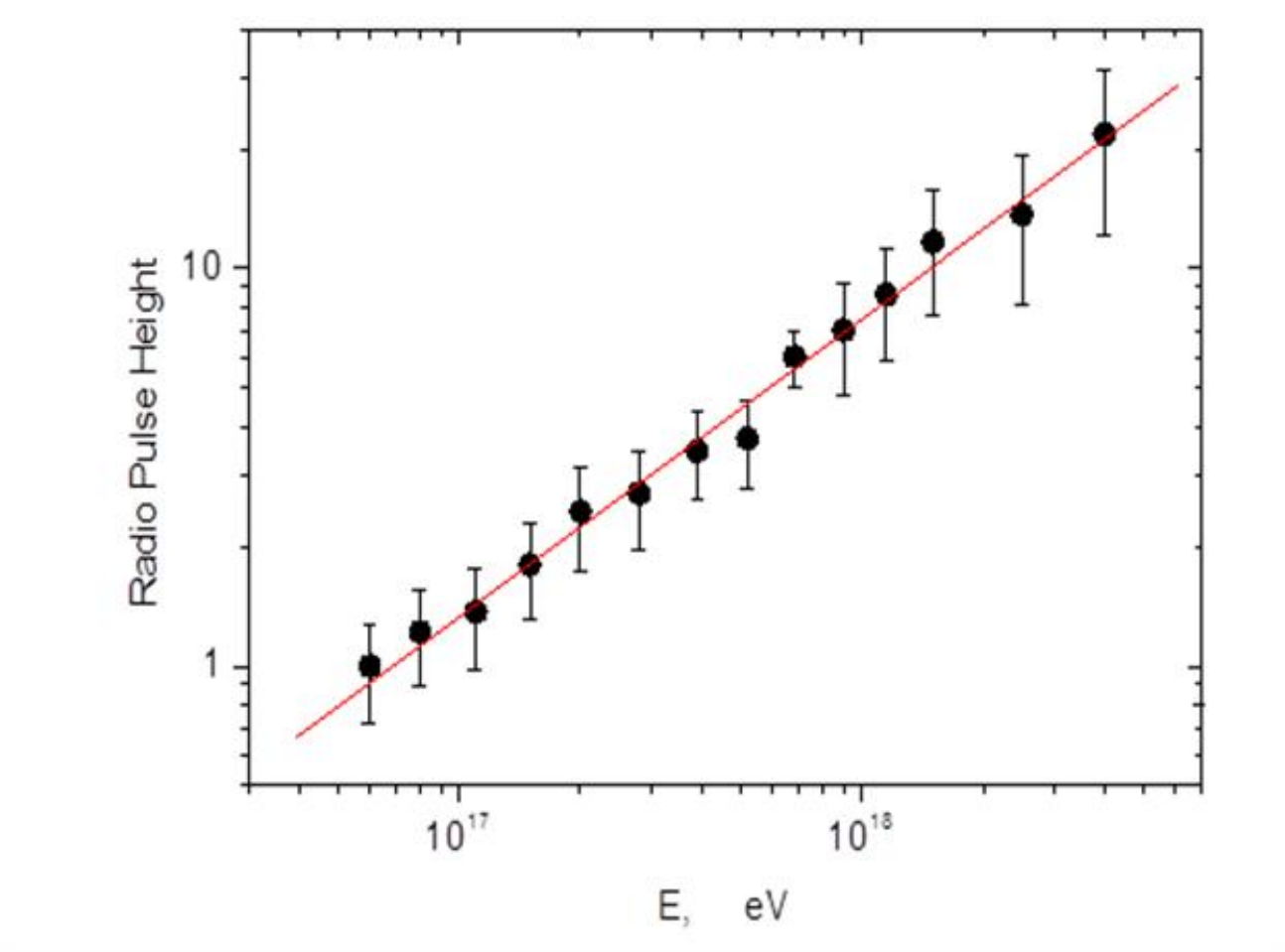}
  \caption{Dependence of maximum amplitude from shower energy.}
  \label{icrc2013-0181-02}
\end{figure}

In Fig. $\ref{icrc2013-0181-02}$ is shown dependence of maximum amplitude of radio signal from shower energy.

Approximation is given by:

\begin{equation}
    \varepsilon(E) = (1.3\pm0.3)(E_{0}/10^{17} eV)^{0.99\pm0.04}
\end{equation}

As seen from Fig. $\ref{icrc2013-0181-03}$, slope of the LDF changes with the distance. At large distance signal of radio emission attenuates slowly. From equation (1) and (2) we derived formula for calculating energy in individual showers:

\begin{figure}[t]
  \centering
  \includegraphics[width=0.4\textwidth]{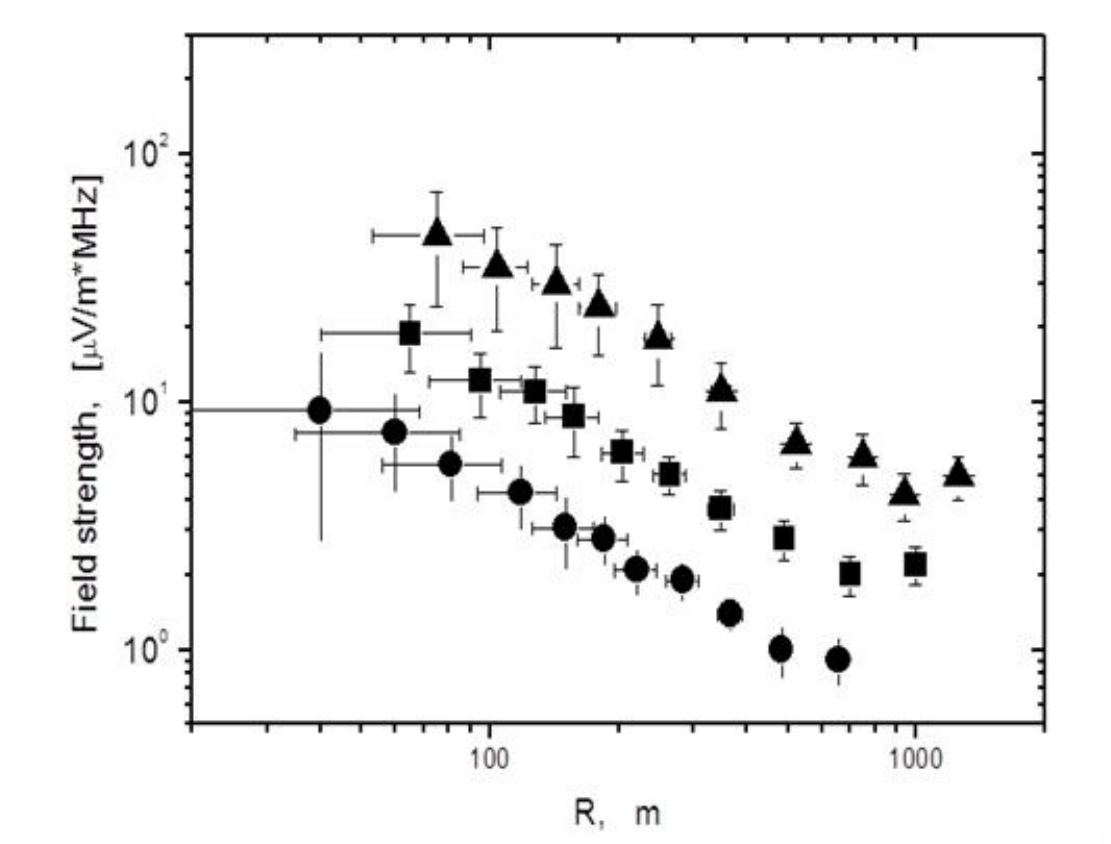}
  \caption{Average lateral distribution function of radio emission at frequency 32 MHz in showers with energy
  1.73$\cdot$10$^{17}$ eV, 4.38$\cdot$10$^{17}$ eV and 1.32$\cdot$10$^{18}$ eV.}
  \label{icrc2013-0181-03}
\end{figure}

\begin{eqnarray}
    \varepsilon (E,\theta, R) = (15\pm1)(1-\cos\theta)^{1.16\pm0.05}\nonumber \\
    \exp\left(-\frac{R}{350\pm25.41}\right)
    \cdot(E_{0}/10^{17} eV)^{0.99\pm0.04}
\end{eqnarray}

With: $\theta$ - zenith angle, R - distance of antennas to the shower axis, E$_{0}$ - the primary particle energy.

\begin{figure}[t]
  \centering
  \includegraphics[width=0.4\textwidth]{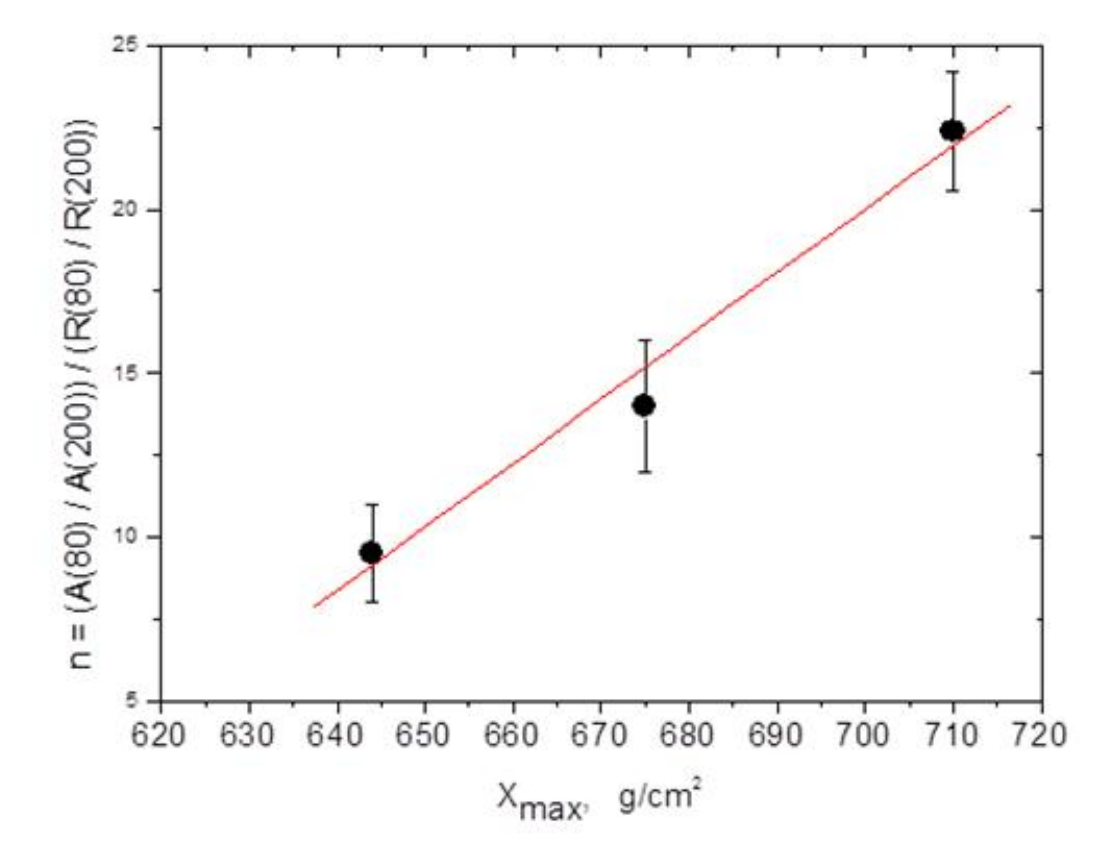}
  \caption{Dependence of air shower radio emission LDF slope n from depth of air shower maximum X$_{max}$. The slope was determined from the ratio of the amplitudes, taken at 80 and 200 m using averaged LDF and Xmax from Cherenkov light measurements. }
  \label{icrc2013-0181-04}
\end{figure}

In Fig. $\ref{icrc2013-0181-04}$ dependence of radio emission LDF from depth of the air shower maximum. The depth was determined by Cherenkov detectors measurements of Yakutsk array. The slopes were determined from the ratio of the amplitudes, taken at 80 and 200 m (Fig. $\ref{icrc2013-0181-03}$) for three energies: 1.73$\cdot$10$^{17}$ eV, 4.38$\cdot$10$^{17}$ eV and 1.32$\cdot$10$^{18}$ eV.

\section{Conclusion}

Yakutsk array measurements are showing: a) there is a correlation between measured maximum amplitude of radio signal and air shower energy determined measurements of the main components at observation level. It follows from the formula (2) derived empirically from the joint consideration of radio signal amplitude and EAS energy; b) the shape of LDF depends on the depth of air shower maximum X$_{max}$ (Fig. $\ref{icrc2013-0181-04}$).

{\bf Acknowledgements:} We express our thanks to the Ministry of Education and Science of the Russian Federation for the financial support of this study as part of the RFBR 12-02-31442 mol$\_$a.

\end{document}